\documentclass[aps,prl,twocolumn,superscriptaddress,showpacs,amssymb,amsmath]{revtex4-2}  
\preprint{APS/123-QED}
\usepackage{graphicx}% Include figure files
\usepackage{dcolumn}% Align table columns on decimal point
\usepackage{bm}% bold math

\usepackage{amsmath}
\usepackage{appendix}
\usepackage{xcolor}
\usepackage{tikz}

\usepackage{braket}
\usepackage{physics}
\usepackage[caption=false]{subfig}\usepackage{mwe}
\usepackage{blochsphere}
\usepackage{array}
\usepackage[thinlines]{easytable}

\usepackage{color}

\definecolor{orange}{rgb}{0.8, 0.3, 0}
\definecolor{blueviolet}{rgb}{0.2, 0.2, 0.6}

\usepackage[pdftex, % FOR PDFLATEX ONLY
bookmarks=true, % Bookmark bar
colorlinks=true,
allcolors=blueviolet,
pdfstartview={FitH}, % FitBH
]{hyperref}

\newcommand{\dg}{^\dagger}

\newcommand{\oa}{{a}}
\newcommand{\oad}{{a}^{\dg}}
\newcommand{\ob}{{b}}
\newcommand{\obd}{{b}^{\dg}}
\newcommand{\oS}{\textbf{S}}
\newcommand{\oI}{\textbf{I}}
\newcommand{\oF}{\textbf{F}}

\newcommand{\gc}{g_\mathrm{3WM}}

\newcommand{\kai}{\kappa_\mathrm{i}^\mathrm{A} }
\newcommand{\kac}{\kappa_\mathrm{c}^\mathrm{A} }

\newcommand{\ka}{\kappa_\mathrm{A}}

\newcommand{\kbi}{\kappa_\mathrm{i}^\mathrm{B} }
\newcommand{\kbc}{\kappa_\mathrm{c}^\mathrm{B} }

\newcommand{\kb}{\kappa_\mathrm{B}}

\newcommand{\Pp}{P_\mathrm{3WM}}

\newcommand{\fA}{f_\mathrm{A} }
\newcommand{\fB}{f_\mathrm{B} }
\newcommand{\fp}{f_\text{3WM} }
\newcommand{\fNMR}{f_\text{NMR} }

\newcommand{\IA}{I_\mathrm{A}}
\newcommand{\IB}{I_\mathrm{B}}

\newcommand{\Bz}{B_\mathrm{z}}

\usepackage{cleveref}
\crefrangelabelformat{equation}{(#3#1#4--#5#2#6)}
\crefname{equation}{Eq.}{Eqs.}
\Crefname{equation}{Equation}{Equations}

\usepackage{comment}

\begin{document}
\title{Addressing spins at their clock transition with a frequency- and bandwidth-tunable superconducting resonator}% Force line breaks with \\

\author{Yutian Wen}
\altaffiliation{Present address: Department of Physics and Astronomy, University of Notre Dame, IN, USA.}
\email{ywen2@nd.edu}
\affiliation{
 Quantronics group, Université Paris-Saclay, CEA, CNRS, SPEC, 91191 Gif-sur-Yvette Cedex, France
 }
 \affiliation{
  École Normale Supérieure de Lyon, CNRS, Laboratoire de Physique, F-69342 Lyon, France
 }
\author{V. Ranjan}
 \affiliation{
 Tata Institute of Fundamental Research Hyderabad, 500046 Hyderabad, India
 }
\author{T. Lorriaux}
 \affiliation{
  École Normale Supérieure de Lyon, CNRS, Laboratoire de Physique, F-69342 Lyon, France
 }
\author{D. Vion}
 \affiliation{
 Quantronics group, Université Paris-Saclay, CEA, CNRS, SPEC, 91191 Gif-sur-Yvette Cedex, France
 }
\author{B. Huard}
 \affiliation{
  École Normale Supérieure de Lyon, CNRS, Laboratoire de Physique, F-69342 Lyon, France
 }
\author{A. Bienfait}
 \affiliation{
  École Normale Supérieure de Lyon, CNRS, Laboratoire de Physique, F-69342 Lyon, France
 }
\author{E. Flurin}
 \affiliation{
 Quantronics group, Université Paris-Saclay, CEA, CNRS, SPEC, 91191 Gif-sur-Yvette Cedex, France
 }
\author{P. Bertet}
 \email{patrice.bertet@cea.fr}
\affiliation{
 Quantronics group, Université Paris-Saclay, CEA, CNRS, SPEC, 91191 Gif-sur-Yvette Cedex, France
}

\date{\today}

\begin{abstract}
Solid-state spin ensembles addressed via superconducting circuits are promising candidates for quantum memory applications, offering multimodal storage capability and second-long coherence times at their clock transition. 
Implementing practical memory schemes requires dynamic control over both the resonator frequency and bandwidth. 
In this letter, we report measurements of a superconducting resonator whose frequency can be tuned by passing a DC current through the high-kinetic-inductance thin film, and whose bandwidth can be tuned by parametric coupling to a low-Q buffer resonator. Using this resonator, we address an ensemble of bismuth donors at their clock transition, measuring a Hahn-echo coherence time of 450 ms. We demonstrate RF driving of the bismuth donor hyperfine transitions, as well as dynamic bandwidth control of the resonator. 

\end{abstract}

\maketitle

Despite continuous improvements, the limited coherence time remains a key challenge for superconducting quantum processors, constraining both noisy intermediate-scale quantum (NISQ) algorithms~\cite{preskill_quantum_2018} and quantum error correction schemes~\cite{fowler_surface_2012}. 
A promising solution~\cite{gouzien_factoring_2021} relies on dedicated quantum memory units capable of parallel storage of multiple qubit states over extended periods. 
Amongst various platforms explored to this end~\cite{craiciu_multifunctional_2021, businger_optical_2020, hann_hardware-efficient_2019, zhong_nanophotonic_2017, freer_single-atom_2017, ranjan_multimode_2020}, 
solid-state spin ensembles stand out for their multimodal storage capability~\cite{schuster_high-cooperativity_2010, grezes_towards_2016, probst_inductive-detection_2017, ranjan_multimode_2020, osullivan_random-access_2022} and second-long coherence times~\cite{wolfowicz_atomic_2013}. 
An example of architecture consists in encoding the qubit states in single- or few-photon microwave pulses to be stored in and retrieved from the spin ensemble memory through refocussing sequences~\cite{afzelius_proposal_2013,julsgaard_dynamical_2012, julsgaard_fundamental_2013,bernad_analytical_2025}. 
One difficulty of this approach is to achieve efficient absorption of the wave packet in the spin ensemble. Even with the use of high-quality-factor resonators to enhance the spin-photon interaction strength, efficient photon absorption requires relatively large spin density, which leads to broad ensemble spin linewidth and reduced spin coherence time due to dipolar interactions~\cite{wolfowicz_decoherence_2012}. 

One possible way to solve this issue is to bias spins at specific magnetic fields where their mean magnetic dipoles vanish, allowing for long coherence times even at large spin concentrations. These so-called clock transitions (CTs), or zero first-order-insensitive (ZEFOZ) points, occur in coupled electron-nuclear spin systems with large hyperfine interactions. Enhanced coherence times at CTs have been observed in donors in silicon~\cite{wolfowicz_atomic_2013}, molecular spins~\cite{shiddiq_enhancing_2016}, and rare-earth-ion-doped crystals~\cite{alexander_coherent_2022,tiranov_sub-second_2025}. 
Because CTs occur only at specific magnetic fields, they require resonators tuned precisely to the corresponding transition frequency. 
While this can be achieved via trial and error with fixed-frequency resonators~\cite{ranjan_multimode_2020}, a more general and scalable approach uses resonators whose frequency can be tuned in-situ and dynamically to the CT frequency. Additionally, several quantum memory protocols~\cite{julsgaard_quantum_2013, bernad_analytical_2025} also require dynamical control of the resonator bandwidth, to avoid maser emission when the spin ensemble is inverted by the control pulses. 

Aluminium-based Josephson devices offer many tools to design resonators with tunable characteristics~\cite{scarlino_situ_2022}, but they are usually incompatible with the application of magnetic fields above $\sim 10\,\mathrm{mT}$ that are necessary to bias the spins. Another mechanism, compatible with large magnetic fields, arises from the quadratic dependence of the kinetic inductance of a superconducting wire, $L_k (I) = L_{k0} + \alpha I^2$ on the current $I$ passing through it. Field-resilient kinetic inductance (KI)-based frequency-tunable resonators have been demonstrated~\cite{annunziata_tunable_2010, vissers_frequency-tunable_2015, wu_broad-spectrum_2025} and used for interfacing spins at $\sim 200-500\,\mathrm{mT}$ fields~\cite{asfaw_multi-frequency_2017}. Here, we design and fabricate a frequency- and bandwidth-tunable resonator based on KI. We demonstrate its suitability as a spin-ensemble quantum memory interface by addressing bismuth donor spins in silicon (Bi:Si) at their CT where we measure extended coherence times, and we perform a spectroscopy of Bi:Si sub-gigahertz transitions by electron nuclear double resonance (ENDOR).

\begin{figure}[ht]
\includegraphics[width=\linewidth, height=0.8\textheight, keepaspectratio]{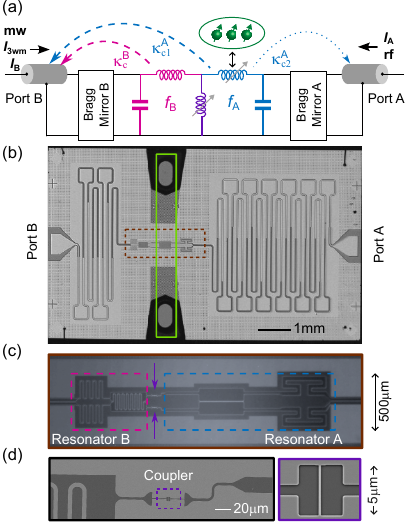}
\caption{\label{fig:device} Device layout. 
(a) Schematic circuit diagram. 
The lumped element models of resonator A (blue) and resonator B (red) are highlighted in (c).
The kinetic inductance coupler (KIC) is shown as an inductive shunt to the ground.
(b) Overview micrograph of the circuit substrate before flip chip bonding. 
The green box outlines the footprint of the Bi-doped silicon die which is subsequently glued on top.
The external polarising field $\Bz$ is applied parallel to the microwire of resonator A, i.e., horizontal in this image.
(c) Zoom-in of the resonator region (brown box in (b)). 
Resonators A and B are marked with dashed boxes.
The arrows point to the KIC.
(d) Optical and scanning electron micrographs of the KIC nanowire. 
The bright areas of various shades are NbTiN films with different vortex-trapping hole densities.
The darkest regions are the exposed silicon substrate. 
}
\end{figure}

The circuit schematics is shown in Fig.~\ref{fig:device}. The core part of the device is a pair of superconducting microwave resonators (``resonator A'' and ``resonator B'' in Fig.~\ref{fig:device}c and the subcircuits of matching colours in Fig.~\ref{fig:device}a) that are inductively coupled via a small mutual inductance, called the kinetic inductance coupler (KIC). 
Resonator A is a high quality-factor resonator which inductively couples to the spin ensemble. It consists of a $800$~nm-wide inductive wire with large KI (referred to as microwire A) and a capacitor in parallel.
Resonator B is a low-$Q$ auxiliary that acts as a tunable microwave buffer for resonator A. Its inductor is $4~\mu \mathrm{m}$ wide. 
The KIC comprises a pair of $5~\mu \mathrm{m}$-long, $250$\,nm-wide nanowires (Fig.~\ref{fig:device}d).
An inductor is tunable when it owes a large fraction of its value to kinetic inductance, as ensured by lateral confinement for microwire A and the KIC.
Microwire B is much wider, and has next to no tunability, as we confirm shortly thereafter.
The resonator frequency $\fA$ can be tuned by passing a DC current $\IA$ through microwire A. 
Its bandwidth tuning is achieved via three-wave mixing with resonator B through the KIC~\cite{vissers_frequency-tunable_2015, malnou_three-wave_2021}.
Both tuning processes require low-frequency or DC bias of the nonlinear inductors.
To this end we flank the dual resonator on either side with notch filters, centred at $\fA$ and $\fB$ respectively.
These so-called ``Bragg mirrors''~\cite{abe_electron_2011}, consisting in successions of quarter-wave-length coplanar waveguides (CPW) of alternating impedances, can maintain high quality factors while permitting DC current and radio-frequency (RF) pump tones to pass through. 
The separate control of DC currents $\IA$ and $\IB$ injected into the devices ports implies independent biases of microwire A ($\IA$) and the KIC ($\IA+\IB$), so that the bandwidth and frequency tunabilities are in principle decoupled.
Note that the Bragg mirror A contains twelve \{high-$Z$, low-$Z$\} repetitions, whereas the Bragg mirror B contains only four. 
As a result, microwave transmission through port A is less than $10^3~\mathrm{s}^{-1}$ and thus negligible, so for both resonators the only relevant microwave coupling are the ones through Bragg mirror B, the rate of which we denote as $\kac$ and $\kbc$ hereinafter.
Accordingly, the internal loss rates are denoted by $\kappa_\mathrm{i}^{\mathrm{A(B)}}$, and the respective total linewidths $\kappa^{\mathrm{A(B)}}=\kappa_\mathrm{i}^{\mathrm{A(B)}}+\kappa_\mathrm{c}^{\mathrm{A(B)}}$.
The whole circuit is patterned on a 50~nm-thick NbTiN film, on which we measured kinetic inductance per square $L_\mathrm{kin}^\mathrm{sq}=2.2~\mathrm{pH/\square}$. 
The circuit is cooled to 10~mK in a dilution refrigerator, with appropriate attenuation and filtering to suppress thermal noise.

\begin{figure}[t!]
\includegraphics[width=\linewidth, height=0.8\textheight, keepaspectratio]{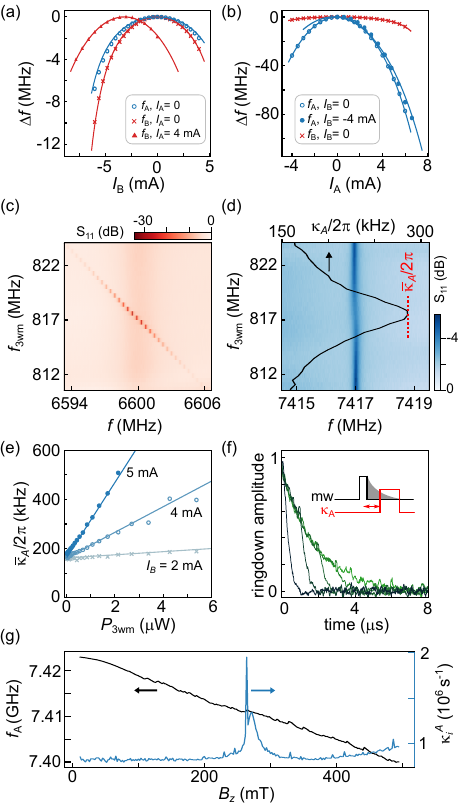}
\caption{\label{fig:basic3wm} 
Resonator tunabilities. 
(a-b) Resonance frequency shift of the two modes in response to sweeping DC bias current $\IA$ (a) or $\IB$ (b), under different $\IB$ or $\IA$ offsets.
Solid curves: fits to Ginzburg-Landau theory~\cite{annunziata_tunable_2010}.
(c-d) Amplitude of the microwave reflection $S_{11}$ off port B near the resonance frequency of mode B (c) and A (d), as the RF pump frequency $\fp$ is swept across the frequency difference $\fA-\fB$.
(e) Dependence of the measured mode A linewidth $\kai$ (open circles) on the RF pump amplitude, under various DC bias through $\IB$ while $\IA$ is held neutral.
(f) Ringdown suppression via dynamic control of resonator bandwidth. 
The green curves of varying shades represent the ringdown amplitude after a resonant microwave pulse, as the 3WM pump pulses are delayed by $0~\mu$s, $1~\mu$s, ..., $4~\mu$s.
The pulse sequences are shown in insets with the varying parameters marked in red with arrows.
(g) Resonance frequency $\fA$ (left axis) and internal loss rate $\kai$ (right axis) of mode A as functions of the magnetic field $\Bz$ applied in-plane, parallel to microwire A. 
}
\end{figure}

We first characterize the circuit by microwave reflectometry through port B. In absence of bias current, we find 
$\fA = 7.422$~GHz, $\kac = 9.4\times10^4$/s, $\kai = 7.5\times10^5$/s;
$\fB = 6.605$~GHz, $\kbc = 2.6\times10^7$/s, $\kbi = 5.7\times10^6$/s.
The resonator frequency shift $\Delta f$ is then measured as functions of $\IA$ and $\IB$ (Fig.~\ref{fig:basic3wm}ab). 
We observe that resonator A can be tuned over an $\sim 80$\,MHz range, whereas resonator B shifts by less than $10$~MHz, as designed. 
All tuning curves fit well to simple Ginzburg-Landau theory~\cite{annunziata_tunable_2010}, from which we can extract the critical current of microwire A, $9.53$~mA, and that of the KIC, $5.73$~mA.
We note the critical currents are proportional to the total cross-section areas, as expected~\cite{van_duzer_principles_1999}.

In addition to frequency tuning, the KIC creates a three-wave mixing (3WM) term $(\oa+\oad)(\ob+\obd)^2$ in the circuit Hamiltonian~\cite{parker_degenerate_2022}, where $\oa$ ($\ob$) is the annihilation operator of mode A (B), and $\oad$ ($\obd$) the creation operator. When a strong drive tone is applied through port B at frequency $\fp = \fA - \fB$, this terms simplifies to a frequency-conversion Hamiltonian $\gc (\oa\obd+\oad\ob)$~\cite{tien_parametric_1958}, with $\gc$ being proportional to the pump amplitude $\sqrt{\Pp}$ and DC current $\IA+\IB$ flowing through the KIC. Since resonator B has a significantly larger decay rate than resonator A, whenever the 3WM is enabled, it translates into an additional loss rate $\kai = 4 \gc^2/\kb$ for resonator A (see Figs.~2c and d)~\cite{Supplemental}. 
The linear dependence of $\kai$ on $\Pp$ and $(\IA+\IB)^2$ is confirmed in Fig.~\ref{fig:basic3wm}e.
The bandwidth can also be tuned dynamically. It is for instance possible to suddenly accelerate the resonator field decay, as demonstrated in Fig.~\ref{fig:basic3wm}f.

We finally measure the resonator A frequency and internal loss rate as a function of a magnetic field $B_z$ applied approximately parallel to the sample surface and to the microwire (Fig.~\ref{fig:basic3wm}f). The frequency decreases due to the kinetic inductance dependence on magnetic field. The loss rate shows an increase around $260$\,mT, and otherwise remains constant except for a slight increase close to $500$~mT. A closer look into the $260$~mT feature reveals a narrow peak that we tentatively ascribe to dangling bond spins at the Si/SiO$_2$ interface with $g=2.0$ (so-called Pb0 centers~\cite{poindexter_interface_1981}), and a broader peak centred at $g= 1.922$, similar to the one called USO in~\cite{bahr_improving_2024} and that has been ascribed to a Ti-related paramagnetic defect.

\begin{figure}[t!]
\includegraphics[width=\linewidth, height=0.8\textheight, keepaspectratio]{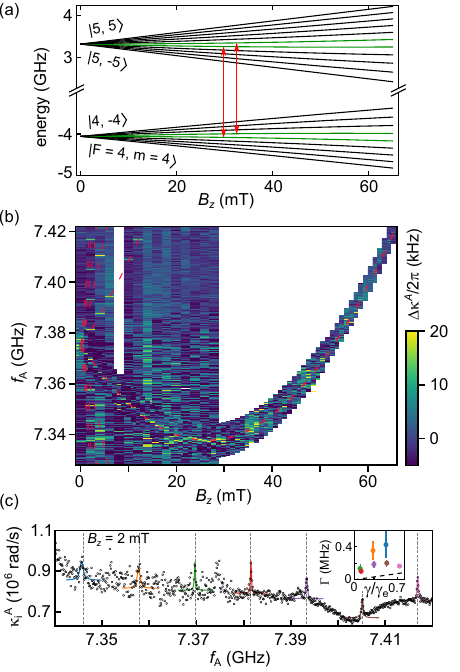}
\caption{\label{fig:cwESR} 
Tracking bismuth spin transitions with a tunable resonator. 
(a) Calculated bismuth donor energy spectrum as a function of the polarising field $B_z$~\cite{mohammady_bismuth_2010, morley_initialization_2010}. 
The arrows mark the clock transitions characterised in Fig.~\ref{fig:pESR}, and the relevant spin states are coloured green.
(b) The resonator absorption spectrometry, measured by the internal loss rate $\kai$ of mode A as its resonance frequency $f$ is swept across the tuning range, and $\Bz$ between 0 and 65~mT. 
The contrast is enhanced and the field-independent background is removed for visibility.
The red dashed curves indicate the calculated transition frequencies.
(c) The absorption spectrum (open circles) at $2.1$~mT. The bismuth transition peaks are indicated by the gray dash lines. 
Inset: the extracted bismuth transition peak width $\Gamma$ versus the gyromagnetic ratio $\gamma$ in unit of the Bohr magneton $\gamma_\mathrm{e}$. 
The dashed line represents the homogenous linewidth corresponding to an Overhauser field $\delta B_0=4~\mu$T due to the residual 500 ppm of $^{29}$Si~\cite{abe_electron_2011, george_electron_2010}.
}
\end{figure}

We now demonstrate that the frequency- and bandwidth-tunable resonator can be used to address spins in solids. A $^{28}\mathrm{Si}$-enriched silicon chip is bonded on top of microwire A, the surface layer facing the resonator pre-implanted with $^{209}\mathrm{Bi}^+$ ions. 
Bismuth is a donor for silicon, which resumes a neutral state at low temperatures. Its spin Hamiltonian is 
$H=(\gamma_e\oS+\gamma_n\oI)B_z+A\oI\cdot\oS$, 
with $\oS$ ($\oI$) being the electron (nuclear) spin operator, 
$\gamma_e/2\pi = - 28$\,GHz/T ($\gamma_n/2\pi = 8$\,MHz/T) the electron (nuclear) spin gyromagnetic ratio, 
and $A/2\pi = 1.47507$\,GHz the hyperfine coupling constant~\cite{A_difference}. 
At $B_z = 0$, the energy states are eigenstates of the total angular momentum operator, $\oF=\oI+\oS$, with eigenvalue $F=4$ for the ground-state manifold of $9$ levels, and $F=5$ for the excited state manifold of $11$ levels. Application of a small magnetic field lifts the degeneracy between levels with different $m$ values of the $F$ projection on $z$ (see Fig.~\ref{fig:cwESR}a). An oscillating magnetic field perpendicular to $B_z$ can induce transitions between these levels whenever $\Delta m = \pm 1$. This can occur at microwave frequency for transitions between levels with $F=4$ and $F=5$ (electron spin resonance (ESR)–like transitions), or at radio-frequency between levels within the same $F$ manifold (nuclear magnetic resonance (NMR)–like transitions). Owing to this peculiar energy-level diagram, several CTs can be found in the Bi:Si spectrum~\cite{mohammady_bismuth_2010, wolfowicz_atomic_2013}. Here we will concentrate primarily on the one at $25.6$\,mT and $7.3382$\,GHz~\cite{A_difference}. 

We perform the spin spectroscopy, first by measuring the resonator loss rate as a function of frequency $\fA$ (by application of $\IA$) and polarising field $B_z$. 
We subtract the field-independent background and plot the change in internal loss rate $\Delta\ka$ in Fig.~\ref{fig:cwESR}(b). 
$B_z$-dependent resonant losses are observed, and their position matches the ESR-like transitions of Bi:Si. 
The $B_z = 2.1$\,mT data are plotted in Fig.~\ref{fig:cwESR}c; the Bi:Si resonances have linewidth of $\sim 300$~kHz, indicative of significantly lower strain than in devices where the metallic resonator was deposited directly on top of the crystal~\cite{bienfait_reaching_2016, pla_strain-induced_2018, ranjan_spatially_2021}, but they are not yet at the homogenous linewidth limit. 
We remove part of the spectrum around $B_z=9$\,mT where aluminium bonding wires were transiting from superconducting to normal states.

\begin{figure}[t!]
\includegraphics[width=\linewidth, height=0.8\textheight, keepaspectratio]{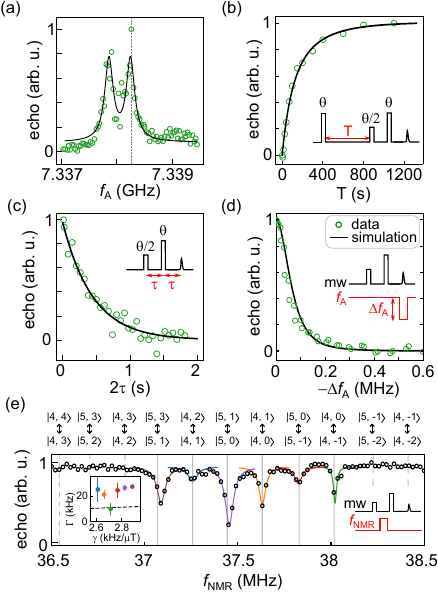}
\caption{\label{fig:pESR} Clock transitions and electron nuclear double resonance (ENDOR). 
The open circles represents the raw data, and the solid curves Lorentzian (a, e) or exponential (b-d) fits. 
The pulse sequences are shown in insets with the varying parameters marked in red with arrows.
(a) Hahn echo spectroscopy at $25.6$~mT. 
$T_1$ (b) and $T_2$ (c) measurement at 7.3382~GHz (dashed grey line in (a)).
(d) Echo silencing via resonator shifting.
The normalised echo magnitude decays with increasing resonance frequency detuning $\Delta\fA$ of mode A during the period of echo.
(e) Normalised microwave echo magnitude (open circles) as a function of the disruptive radio frequency pulse frequency $\fNMR$. 
The grey lines indicate the nuclear magnetic resonance (NMR)–like transitions. The ones involving $\ket{4,1}, \ket{4,0}, \ket{5, 0}$, or $\ket{5, 1}$ are highlighted with solid lines.
Left inset: the extracted peak width $\Gamma$ of the NMR-like transitions versus their gyromagnetic ratio $\gamma$. 
The dashed line represents the homogenous linewidth corresponding to a magnetic noise $\delta B_0=4~\mu$T.
}
\end{figure}

The Bi:Si spectrum at the CT is measured by sweeping the resonator frequency $\fA$ and recording the amplitude of a Hahn echo. It consists of two resolved peaks of width $\sim 90$\,kHz, corresponding respectively to the transitions $\ket{4, -1}\leftrightarrow \ket{5, 0}$ and $\ket{4, 0}\leftrightarrow \ket{5, -1}$ (see Fig.~\ref{fig:cwESR}a). The energy relaxation time is measured to be $T_1=53$~s by an inversion recovery sequence (see Fig.~\ref{fig:pESR}a). This is two orders of magnitude shorter than typical non-radiative relaxation times in Bi:Si at $10$\,mK, indicating that the donor spins are well in the Purcell regime~\cite{bienfait_controlling_2016,albanese_radiative_2020}. We then measure the echo amplitude as a function of the Hahn echo delay $\tau$ (see Fig.~\ref{fig:pESR}b). The data are well-fitted by an exponential decay, with a time constant $T_2 = 450$\,ms, similar to those already observed at Bi:Si CTs using fixed-frequency resonators~\cite{wolfowicz_atomic_2013, ranjan_multimode_2020}. This demonstrates that we can address the Bi:Si donor spins at their CT, using a frequency-tunable resonator. 
Echo emission can be controlled by dynamically tuning the resonator away from resonance at the time of the echo formation, as already demonstrated in Ref.~\cite{ranjan_spin-echo_2022, de_graaf_scaling_2024}. Such echo silencing is useful in several quantum memory protocols~\cite{damon_revival_2011,afzelius_proposal_2013,julsgaard_quantum_2013}.

The ability to send sub-gigahertz pulses is harnessed to drive NMR-like transitions between Bi:Si levels. 
To that goal, the amplitude of a Hahn echo is measured as a function of the frequency $f_\mathrm{NMR}$ of a radio frequency pulse sent in-between the two echo control pulses.
With the magnetic field set to $\Bz=13.49$~mT and the resonator to 7422.5~MHz, we primarily probe the $\ket{4,0}\leftrightarrow\ket{5,1}$ transition, though the $\ket{4,1}\leftrightarrow\ket{5,0}$ transition also contributes slightly due to spectral overlap. The echo signal diminishes when $\fNMR$ matches an NMR–like transition involving one of these states.
This is evident in Fig.~\ref{fig:pESR}(e), where the dips on the normalised echo magnitude correspond to the relevant NMR-like transition spectrum, from low to high frequency:
$
\ket{5,2}\leftrightarrow\ket{5,1},
\ket{4,2}\leftrightarrow\ket{4,1},
\ket{5,1}\leftrightarrow\ket{5,0},
\ket{4,1}\leftrightarrow\ket{4,0},
\ket{5,0}\leftrightarrow\ket{5,-1},
\ket{4,0}\leftrightarrow\ket{4,-1}
$.
In particular, the 
$
\ket{4,2}\leftrightarrow\ket{4,1},
\ket{5,0}\leftrightarrow\ket{5,-1}
$
dips are visibly shallower than the other four, as expected from the lower contribution of levels $\ket{4,1}$ and $\ket{5,0}$ to the echo signal. Other NMR-like transitions have no impact on the echo amplitude.

In conclusion, we have demonstrated a field-resilient resonator with tunable frequency and bandwidth, suitable for addressing paramagnetic impurities. Such device should find applications in microwave quantum memories, as well as nanoscale magnetic resonance spectroscopy~\cite{tyryshkin_electron_2012, sigillito_fast_2014, rose_coherent_2017}.

\begin{acknowledgments}
The authors acknowledge the technical support of P. Sénat, D. Duet, P.-F. Orfila, and S. Delprat, and fruitful discussions within the Quantronics group.
The authors acknowledge the support of the AIDAS joint laboratory,
of Région Ile-de-France through the DIM SIRTEQ, 
of the Agence Nationale de la Recherche under the Chaire Industrielle NASNIQ, 
and under the PEPR Plan Project ROBUSTSUPERQ, 
and IARPA and Lincoln Labs for providing the Josephson travelling-wave parametric amplifier.
\end{acknowledgments}

\onecolumngrid

\appendix
\renewcommand{\thefigure}{S\arabic{figure}}
\setcounter{figure}{0}
\renewcommand{\theequation}{S\arabic{equation}}
\setcounter{equation}{0}

\section{Supplemental materials to ``Addressing spins at the clock transitions with a frequency- and bandwidth-tunable superconducting resonator''}
\subsection{Sample}

The bottom chip containing the superconducting device is made out of 50-nm-thick NbTiN deposited on Silicon. 
Our fabrication starts from a 2" Si wafer (high-resistivity grade supplied from Siltronics), that is deoxidised in 5\% HF for two minutes before NbTiN sputtering. 
The device fabrication process comprises three lithography steps, using either a 30 keV Raith electron-beam lithography system or a Heidelberg Instruments Maskless Aligner. 
Afterwards the pattern is either converted to 30~nm of aluminium hard mask via lift-off, or we use directly the developed photoresist as a soft mask for reactive ion etching to remove the appropriate parts of the NbTiN film or the silicon substrate.

The cover chip is a silicon chip topped by a 800~nm-thick epilayer of $^{28}$Si. Bismuth atoms have been implanted in this layer with the following implantation energies and fluences: 

\begin{center}
\begin{tabular}{cc}
\hline
\textbf{Ion Energy (keV)} & \textbf{Area Dose (Ions/cm$^2$)} \\
\hline
2300 & $1.585 \times 10^{12}$ \\
2000 & $1.2184 \times 10^{12}$ \\
1400 & $8.9772 \times 10^{11}$ \\
900  & $6.6804 \times 10^{11}$ \\
500  & $3.1748 \times 10^{11}$ \\
360  & $1.0517 \times 10^{11}$ \\
120  & $7.0692 \times 10^{10}$ \\
\hline
\textbf{Total Dose} & $4.8626 \times 10^{12}$ \\
\hline
\end{tabular}
\end{center}

\begin{figure}[ht]
     \centering
     \includegraphics[width=0.5\linewidth]{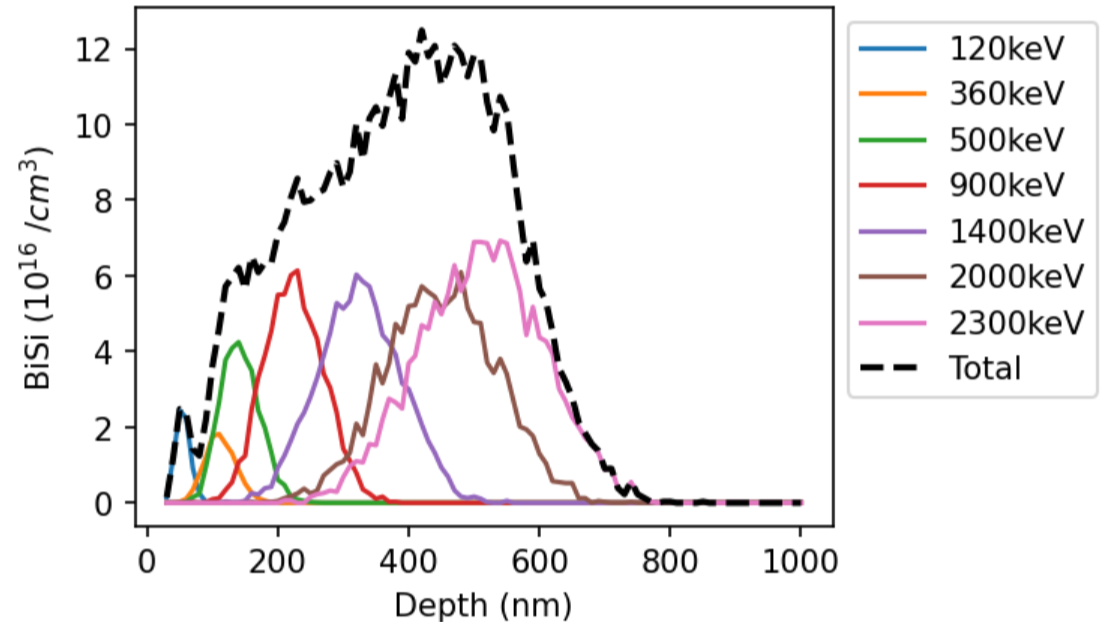}
     \caption{Stopping and Range of Ions in Matter (SRIM) simulation of the bismuth donor concentration as a function of the distance from the surface.}
     \label{fig:SRIMS}
 \end{figure}

We can then estimate the implantation profile using Stopping and Range of Ions in Matter (SRIM) simulation, as shown in Fig.~\ref{fig:SRIMS}. 

The two chips have been assembled in a flip-chip bonding machine, glued together by PMMA 950A6 only at the contact pillars of the base chip (elliptical structures at the ends of the green box in Fig.~\ref{fig:device}b).
Two 200~$\mu$m-deep reservoirs are etched out around the pillars (dark pentagon regions) to contain excess PMMA, preventing it from wetting the entire device region through capillary action.

\subsection{Measurement setup}

\begin{figure}
     \centering
     \includegraphics[width=1\linewidth]{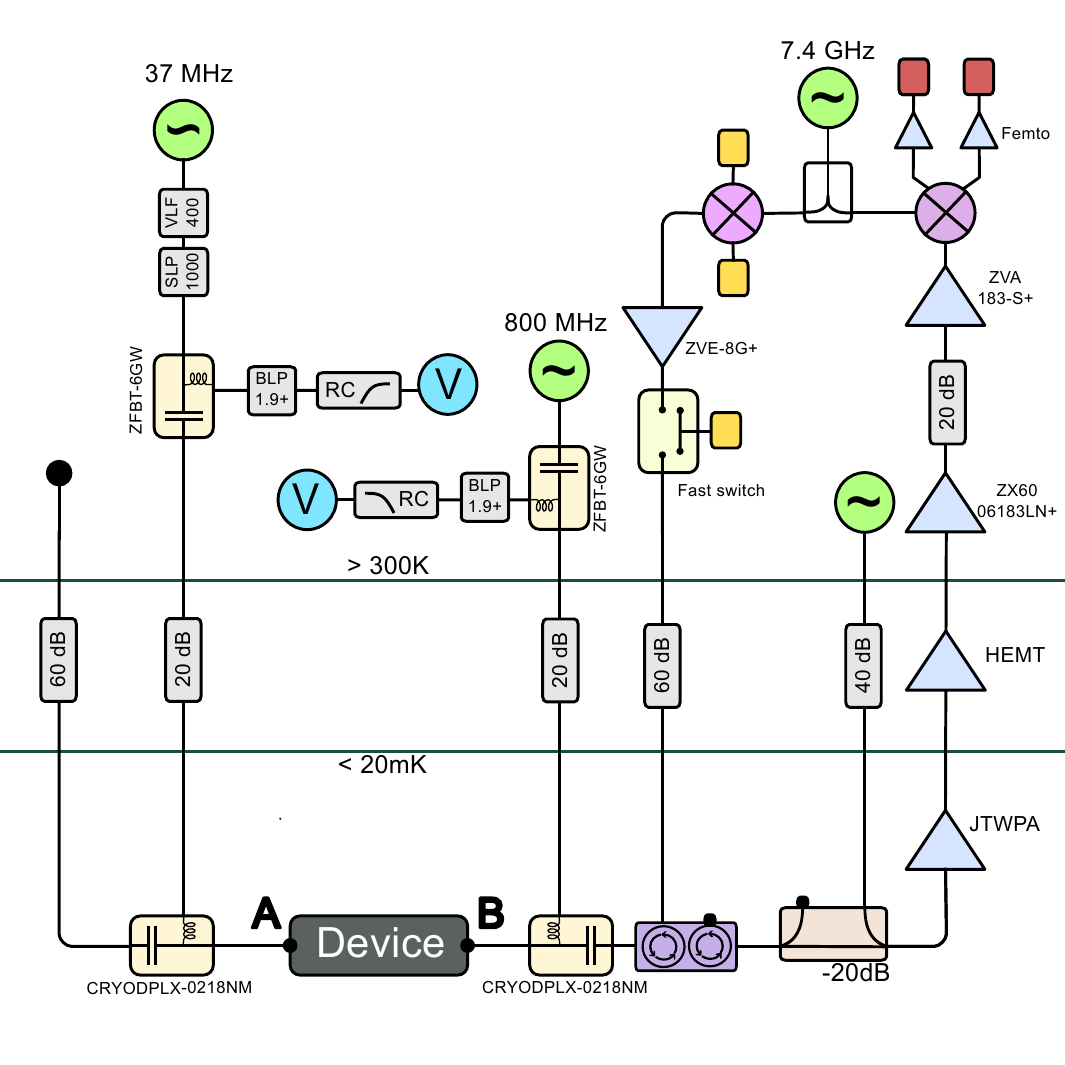}
     \caption{Experimental setup at room temperature and at low-temperature. The microwave tone driving the spins is controlled through IQ-mixing as well as with fast switches. Its transmission, or any signal emitted from the spins, is amplified by a series of four amplifiers: a Josephson parametric amlifier provided by the Lincoln lab, a HEMT amplifier, and two mini-circuits amplifiers. The DC currents controlling the KIC and microwire inductances are generated by Yokogawa voltage sources. Both channels are fitted with home-made RC filters ($225~\Omega$, 220~pF). On the buffer side, the dc current is combined with the 3 wave-mixing tone using a bias-tee at room-temperature, and is further recombined with the microwave signal at the mixing-chamber stage. 
    On the other side, the current is recombined at low temperature with the rf drive for nuclear spin manipulation. An Anapico APSUASYN30-4 generates the 4 microwaves tones necessary to the experiments (rf, three-wave mixing, JTWPA pump, and microwave drive tone). A Quantum Machine OPX generates the low-frequency control signals (yellow squares) and perform the acquisition of the low-frequency signals (red squares). The experiment is realized in a Bluefors LD system equipped with a 3D vector magnet.
    Some unused components are omitted for simplicity.}
     \label{fig:setup}
 \end{figure}

The cryogenic and room temperature setups are shown in Fig.~\ref{fig:setup}.

\subsection{Dissipation engineering through three-wave mixing using kinetic inductance non-linearity}

\subsubsection{Kinetic inductance modulation}
In this section, we detail the action of the three-wave mixing drive on the coupled resonators system. This drive induces a modulation of the kinetic inductance of the microwire and of the nanowire. The frequencies of resonators $A$ and $B$ show a quartic dependence on their biasing current (see Fig.~2(a-b)), so that the kinetic inductance dependence on current can be modelled in all generality as
\begin{equation}
    L_k(I)= L_k^0 [ 1 + \left(\frac{I}{I^*}\right)^2 + \alpha \left(\frac{I}{I^*}\right)^4],
\end{equation}
where $\alpha \lesssim 1$ can be determined by fitting the experimental data. 
When applying a rf current drive $I_{\text{rf}}$ at fixed bias $I_{\text{dc}}$, the inductance modulation thus expresses as 
\begin{equation}
    L(I_{\text{dc}}+I_{\text{rf}})= L_g + L_k(I_{\text{dc}}) + L_k(0)\sum_{i=1}^{4} c_i I_{\text{rf}}^i,
\end{equation}
where we have included the geometrical contribution $L_g$ and where:
\begin{align}
    c_1/I^* &= 2 \frac{I_{\text{dc}}}{I^*} + 4 \alpha\left(\frac{I_{\text{dc}}}{I^*}\right)^3,\\
    c_2/I^{*2} &= 1 + 6 \alpha \left(\frac{I_{\text{dc}}}{I^*}\right)^2 ,\\
    c_3/I^{*3} &= 4\alpha\left(\frac{I_{\text{dc}}}{I^*}\right),\\
    c_4/I^{*4} &= \alpha
\end{align}
For small drive currents ($I_{\text{rf}}\ll I^*$) and at small finite bias ($I_{\text{dc}}<I^*)$, one can safely approximate the kinetic inductance modulation up to its second order expansion, i.e. discarding all terms in $\alpha$.

\subsubsection{Circuit Hamiltonian}
 \begin{figure}[ht]
     \centering
     \includegraphics[width=0.5\linewidth]{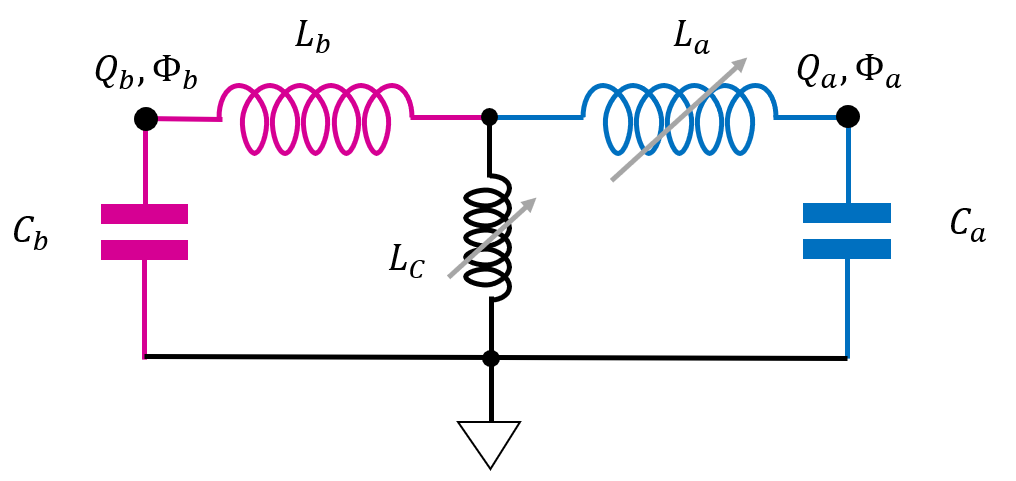}
     \caption{Device circuit diagram.}
     \label{fig:qcircuit}
 \end{figure}
We now wish to diagonalize the circuit (Fig.~\ref{fig:qcircuit}), taking into account the kinetic inductance non-linearity. Denoting $C_{\mathrm{a}}$ and $C_{\mathrm{b}}$ ($L_{\mathrm{a}}$ and $L_{\mathrm{b}}$) the capacitance (inductance) of each resonator, and $L_{\mathrm{c}}$ the coupling KIC inductor, the Lagrangian of the circuit is~\cite{yurke_quantum_1984}:
\begin{align}
    \mathcal{L} = - \frac{Q_{\mathrm{a}}^2}{2 C_{\mathrm{a}}} - \frac{Q_{\mathrm{b}}^2}{2 C_{\mathrm{b}}} + \frac{L_{\mathrm{a}}}{2} \dot{Q}_{\mathrm{a}}^2 + \frac{L_{\mathrm{b}}}{2} \dot{Q}_{\mathrm{b}}^2 + \frac{L_{\mathrm{c}}}{2} (\dot{Q}_{\mathrm{a}} + \dot{Q}_{\mathrm{b}})^2,
\end{align}
where $Q_{\mathrm{a}}$ and $Q_{\mathrm{b}}$ are the charge on each capacitor. We consider that only $L_{\mathrm{a}}$ and $L_{\mathrm{c}}$ are non-linear inductors, so that their values depend on the currents flowing through them (respectively $\dot{Q}_{\mathrm{a}}$ and $\dot{Q}_{\mathrm{a}}+\dot{Q}_{\mathrm{c}}$) as described above.
\begin{align}
    L_{\mathrm{a}} &= L_{\mathrm{a}}^0 + c_1^a \dot{Q}_{\mathrm{a}} + c_2^a \dot{Q}_{\mathrm{a}}^2 \\
    L_{\mathrm{c}} &= L_{\mathrm{c}}^0 + c_1^n (\dot{Q}_{\mathrm{a}}+\dot{Q}_{\mathrm{b}}) + c_2^n (\dot{Q}_{\mathrm{a}}+\dot{Q}_{\mathrm{b}})^2.
\end{align}

The Hamiltonian of the circuit is given by:
\begin{equation}
    H = \Phi_{\mathrm{a}} \dot{Q}_{\mathrm{a}}+ \Phi_{\mathrm{b}} \dot{Q}_{\mathrm{b}} - \mathcal{L},
\end{equation}
where the phase operators can be evaluated through $\Phi_{i}=\frac{\partial  \mathcal{L}}{\partial Q_{i}}$, yielding:
\begin{align}
    \Phi_{a} &= (L_{\mathrm{a}}^0 + L_{\mathrm{c}}^0) \dot{Q_{\mathrm{a}}} + L_{\mathrm{c}}^0 \dot{Q_{\mathrm{b}}} + \frac{3}{2} c_1^a \dot{Q}_{\mathrm{a}}^2 + \frac{3}{2} c_1^n  (\dot{Q}_{\mathrm{a}}+\dot{Q}_{\mathrm{b}})^2 + 2 c_2^a \dot{Q}_{\mathrm{a}}^3 +  2 c_2^n (\dot{Q}_{\mathrm{a}}+\dot{Q}_{\mathrm{b}})^3 \\
    \Phi_{b} &= (L_{\mathrm{b}} + L_{\mathrm{c}}^0) \dot{Q}_{\mathrm{b}} +  L_{\mathrm{c}}^0 \dot{Q_{\mathrm{a}}} + \frac{3}{2} c_1^n  (\dot{Q}_{\mathrm{a}}+\dot{Q}_{\mathrm{b}})^2  +  2 c_2^n (\dot{Q}_{\mathrm{a}}+\dot{Q}_{\mathrm{b}})^3.
    \label{eq:phidef}
\end{align}
Expressing the Hamiltonian purely as a function of ($\Phi_{\mathrm{a}}$, $\Phi_{\mathrm{b}}$, $Q_{\mathrm{a}}$, $Q_{\mathrm{b}}$) requires inverting Eq.~\ref{eq:phidef} to express $\dot{Q}_{a/b}$ as a function of $\Phi_{\mathrm{a}}$ and $\Phi_{\mathrm{b}}$. We do so by defining the vectors $\Phi = (\Phi_{\mathrm{a}}, \Phi_{\mathrm{b}})$ and $\dot{Q} = (\dot{Q}_{\mathrm{a}},\dot{Q}_{\mathrm{b}})$, and express Eq.~\ref{eq:phidef} as:
\begin{equation}
    \Phi = \Lambda \dot{Q} + R,
\end{equation} where $\Lambda$ is a 2-by-2 matrix independent of $\dot{Q}_{a/b}$ and $R$ contains the terms in $\dot{Q}_{a/b}^k$ with $k>1$. We now have:
\begin{equation}
    \dot{Q}=\Lambda^{-1}\Phi - \Lambda^{-1} R.
    \label{Qapprox}
\end{equation}
One can then express the higher terms contained in $R$ by substituting them by Eq.~\ref{Qapprox}. This creates a new expression for $\dot{Q}$. Performing this process iteratively enables to express $\dot{Q}_{a/b}$ as a function of $\Phi_{\mathrm{a}}$, $\Phi_{\mathrm{b}}$, and higher order cross-products, and these expressions can then be trimmed at a given order before injection into the Hamiltonian. Cutting-off the resulting expressions at the third order (up to $\Phi_{a/b}^3$), we find :
\begin{equation}
    H = H_0 + g_{1,1} \Phi _{\mathrm{a}} \Phi _{\mathrm{b}}+
    g_{2,1} \Phi _{\mathrm{a}}^2 \Phi _{\mathrm{b}} +g_{1,2} \Phi _{\mathrm{a}} \Phi _{\mathrm{b}}^2+g_{3,0} \Phi _{\mathrm{a}}^3+g_{0,3} \Phi _{\mathrm{b}}^3
    \label{eq:H}
\end{equation}
where:
\begin{align}
    H_0 &=\frac{Q_{\mathrm{a}}^2}{2 C_{\mathrm{a}}} + \frac{Q_{\mathrm{b}}^2}{2 C_{\mathrm{b}}} + \frac{\Phi_{\mathrm{a}}^2}{2 \tilde{L}_{\mathrm{a}}} + \frac{\Phi_{\mathrm{b}}^2}{2 \tilde{L}_{\mathrm{b}}},  \\
    \tilde{L}_{\mathrm{a}} &= L_{\mathrm{a}}^0 + L_{\mathrm{c}}^0 \frac{1}{1+L_{\mathrm{c}}^0/L_{\mathrm{b}}^0}, \\
    \tilde{L}_{\mathrm{b}} &= L_{\mathrm{b}} + L_{\mathrm{c}}^0 \frac{1}{1+L_{\mathrm{c}}^0/L_{\mathrm{a}}}\\
    g_{1,2} &= -\frac{3}{2\tilde{L}_{\mathrm{a}}} c_1^a \left(\frac{ L_{\mathrm{c}}^0}{L_{\mathrm{a}}^0L_{\mathrm{b}}(1+L_{\mathrm{c}}^0/L_{\mathrm{ab}})}\right)^2   -\frac{3}{2} c_1^n \frac{1}{(L_{\mathrm{b}}+L_{\mathrm{c}}^0)^2(L_{\mathrm{a}}^0+L_{\mathrm{c}}^0)}\frac{L_{\mathrm{a}}^0L_{\mathrm{b}}}{\tilde{L}_{\mathrm{a}}\tilde{L}_{\mathrm{b}}}
\end{align}
and others $g_{i,j}$ are geometrical terms that can be expressed as a function of $L_{\mathrm{c}}^0$, $L_{\mathrm{a}}^0$, $L_{\mathrm{b}}$, $c_1^a$, $c_1^n$, but do not involve $c_2^a$ and $c_2^b$. We defined $L_{\mathrm{ab}}=\frac{L_{\mathrm{a}}^0L_{\mathrm{b}}}{L_{\mathrm{a}}^0+L_{\mathrm{b}}}$. We now introduce the annihilation operators $\hat{a} =  (\Phi_{\mathrm{a}}+iZ_{\mathrm{a}}Q_{\mathrm{a}} )/\sqrt{(2\hbar Z_{\mathrm{a}})}$ and $\hat{b} =  (\Phi_{\mathrm{b}}+iZ_{\mathrm{b}} Q_{\mathrm{b}}) /\sqrt{(2\hbar Z_{\mathrm{b}})}$, where $Z_{\mathrm{a}} = \sqrt{\tilde{L}_{\mathrm{a}}/C_{\mathrm{a}}}$  and $Z_{\mathrm{b}} = \sqrt{\tilde{L}_{\mathrm{b}}/C_{\mathrm{b}}}$ are the impedances of the coupled resonators. The Hamiltonian~\ref{eq:H} can be re-expressed as a function of these operators.

\subsubsection{Three-wave mixing for dissipation engineering}

We now wish to derive the effective three-wave Hamiltonian, and show that it can be used to control the dissipation rate of resonator $A$. We thus consider the action of a pump tone of amplitude $\xi$ with frequency $\omega_{\mathrm{p}}\sim\omega_{\mathrm{a}}-\omega_{\mathrm{b}}$ applied on the buffer port, as well as a drive tone applied at $\omega_{\mathrm{d}}\sim\omega_{\mathrm{a}}$. We accomplish this by considering the interacting Hamiltonian $H_{\text{int}} = U H U^\dagger - i \hbar U \frac{d U^\dagger}{dt} $ calculated using the following evolution operator 
\begin{equation}
    U=e^{i\omega_{\mathrm{d}} t \hat{a}^\dagger \hat{a}} e^{i(\omega_{\mathrm{d}}-\omega_{\mathrm{p}}) t \hat{b}^\dagger \hat{b}} e^{-\tilde{\xi} \hat{b}^\dagger + \tilde{\xi}^* \hat{b} }.
\end{equation} 
where we remove the classical displacement due to the application of pump on the buffer flux. The displacement $\tilde{\xi}=\xi e^{-i\omega_{\mathrm{p}} t}$ can be related to the rf drive current flowing through the inductor by $\xi = \tilde{L}_{\mathrm{b}} I_{\text{rf}} /\sqrt{2\hbar Z_{\mathrm{b}}}$. In our experiment, owing to the Bragg mirror, this current can be safely related to the input pump power on the buffer port $P_{\text{in}}$ as $\sqrt{P_{\text{in}}/Z_0}$,  since the pump tone is applied far below the Bragg mirror bandpass frequency range.

Keeping only terms of $H_{\text{int}}$ which are non zero after a time-average over one drive period, we effectively perform the rotating-wave approximation and we find:
\begin{equation}
H_{\text{RWA}} = \hbar \delta_{\mathrm{a}} \hat{a}^\dagger \hat{a}+ \hbar \delta_{\mathrm{b}} \hat{b}^\dagger \hat{b} + \hbar g_{\text{3WM}}  (\hat{a} \hat{b}^\dagger + \hat{a}^\dagger \hat{b})
\label{HRWA}
\end{equation}
where we have $\delta_{\mathrm{a}} = \omega_{\mathrm{a}}-\omega_d$, $\delta_{\mathrm{b}} = \omega_{\mathrm{b}} + \omega_p - \omega_d$ and:
\begin{equation}
    \hbar g_{\text{3WM}} = 4 \xi\sqrt{(2\hbar)^3 Z_aZ_b^2} g_{1,2} = 12 k \frac{I_{\mathrm{dc}}I_{\mathrm{rf}}}{I^{*2}} \hbar\sqrt{\omega_a \omega_b},
\end{equation}
with $k \approx L_{k,n}^0/\sqrt{L_\mathrm{a}L_\mathrm{b}}$ and where we have neglected the contribution of $L_{\mathrm{a}}$ to the third-order non-linearity.

We now derive the quantum Langevin equations from Eq.~\ref{HRWA}, including the dissipation on modes $a$ and $b$, as well as a drive of strength $\alpha_{\text{in}}$ applied to mode $a$:
\begin{align}
\dot{a} &= -\left(\frac{\kappa_{\mathrm{a}}}{2}+ i \delta_{\mathrm{a}}\right) a - i g_{\text{3WM}} b + \sqrt{\kappa_{\mathrm{a}}^\mathrm{c}} \alpha \\
\dot{b} &= -\left(\frac{\kappa_{\mathrm{b}}}{2}+ i \delta_{\mathrm{b}}\right) b - i g_{\text{3WM}}a
\end{align}
In the steady-state, whenever $\delta_\mathrm{b}=0$, we find:
\begin{equation}
    a = \frac{2 \sqrt{\kappa_{\mathrm{a}}^\mathrm{c}} \alpha_{\text{in}}}{\kappa_\mathrm{a} + \frac{4 g^2_{\text{3WM}}}{\kappa_{\mathrm{b}}}+2i\delta_{\mathrm{a}}},
\end{equation}
where we can identify that through the three-wave mixing interaction, the buffer mode creates an additional decay channel for mode $a$ of rate $4 g^2_{\text{3WM}}/\kappa_{\mathrm{b}}$.

\bibliography{Submission_arXiv} 
\bibliographystyle{apsrev4-2}

\end{document}